# A Study of the Large–Scale Distribution of Galaxies in the South Galactic Pole Region: I. The Data *

S. Ettori[1,2], L. Guzzo[1], M. Tarenghi[2]

[1] *Osservatorio Astronomico di Brera, Via Bianchi 46, I-22055 Merate (CO), Italy*
[2] *European Southern Observatory, K. Schwarzschild Str. 2, D-85748 Garching, Germany*

**ABSTRACT**
We present the data from an extensive, moderately deep ($b_J \sim 19.5$) spectroscopic survey of $\sim 600$ galaxies within four regions of sky located near the South Galactic Pole. About 75% of the measured galaxies are in a $\sim 3° \times 1.5°$ region dominated by the rich cluster of galaxies Klemola 44 (Abell 4038). The other three smaller areas cover about 1 square degree each. Here we discuss in detail the observing and data reduction strategies, the completeness and errors on the measured redshifts. The data collected are being used for: (1) a study of the large–scale redshift distribution of the galaxies in each field, and (2) a thorough dynamical investigation of Klemola 44. Results from these analyses will be presented in forthcoming papers.

## 1 INTRODUCTION

During the 1980s the industry of redshift measurements produced some of the most remarkable results on the large–scale structure of the Universe. In particular, the CfA survey (Huchra et al. 1983; Geller & Huchra 1989), and the Perseus–Pisces survey (Giovanelli et al. 1986), covered wide solid angles on the sky, measuring redshifts for complete magnitude–limited subsamples of the Zwicky et al. (1963-68) catalogue. These and other subsequent surveys (see Giovanelli & Haynes 1991 for a review) were all performed with single object spectroscopy, i.e. measuring one spectrum at a time. While the first multiobject spectrographs were being developed during the first half of the decade (Gray 1984; see in particular the two excellent reviews by Ellis and Parry, 1988, and Hill, 1988), the typical surface density of objects at the limiting magnitudes reached by the available catalogues ($m_b \leq 15.5$) was simply too low by a factor of 100 or more for justifying, or even vaguely suggesting, the use of multiplexing in large–scale redshift survey work. To our knowledge, the first use of a multiobject spectrograph for extensive galaxy redshift survey work, although still limited to small fields, was the faint galaxy survey of Broadhurst et al. (1988) using the FOCAP fibre coupler at the AAT.

In 1987 we became aware of the ongoing activity in the UK to construct two large digitized catalogues of galaxies, the Edinburgh/Durham Southern Galaxy Catalogue (EDSGC: Heydon-Dumbleton et al. 1989) and the APM catalogue (Maddox et al. 1990), starting from the same photographic material, i.e. the ESO/SRC J plates. Given their faint limiting magnitude ($b_J \sim 20.5$), these catalogues represented an ideal data base for conceiving a further step in redshift survey work, i.e. using a wide–field multiobject spectrograph for surveying in depth a large area of the sky. For these reasons, in 1988 we started a preliminary program aimed to test the performances of what at the time was a new prototype version of the ESO fibre spectrograph Optopus (Avila et al. 1989). The idea was eventually to start at ESO a dedicated program of large–scale redshift survey down to $b_J \sim 19-19.5$, the magnitude that optimizes the use of the instrument by matching the number of available fibres (Guzzo & Tarenghi, 1987). Here we present the data from this early test survey over four selected regions of sky, which fully confirmed the reliability of Optopus for such a project (see Avila et al. 1989). Eventually, a complete large–scale redshift survey based on the EDSGC to the same limiting magnitude, over a strip of $\sim 30° \times 1°$, was started in 1991 and is presently under completion (ESO Slice Project, ESP hereafter, Vettolani et al. 1994).

Forthcoming papers related to the data presented here will discuss the large–scale redshift distribution in each field and the kinematics of the cluster Klemola 44 / Abell 4038 (Klemola 1969).

---

* Based on data collected at the European Southern Observatory, La Silla, Chile.

## 2 THE SURVEY

### 2.1 Sample Selection

In the planning of a large redshift survey, a basic starting point is represented by the availability of a photometric catalogue of galaxies complete to some limiting magnitude. A notable example in this sense has been represented by the Zwicky et al. (1963–68) *Catalogue of Galaxies and Clusters of Galaxies* (CGCG), which has been used for a number of redshift surveys of optical galaxies (see Oort 1983, Rood 1988, Giovanelli & Haynes 1991 and references therein). At the end of the eighties, however, a major step forward in this field has been achieved through the construction of two large automatic catalogues of galaxies, already mentioned in the Introduction, the EDSGC (Heydon-Dumbleton et al. 1989) and the APM catalogue (Maddox et al. 1990). Both these compilations were constructed by digitizing the plates of the ESO/SRC J photographic survey, and then analyzing the digital data to separate stars from galaxies. Star/galaxy separation is accomplished to better than $\sim 5\%$, and the catalogues are fairly complete to a magnitude $b_J = 20.5$. Since the construction of these catalogues was started with the main goal of using them for cosmological studies, particular care was devoted to the plate-to-plate matching of the internal magnitude system, to avoid that spurious calibration inhomogeneities could mimic large-scale clustering.

The availability to us of parts of the EDSGC since its early stages of construction, represented the driving force which stimulated the observations presented in this paper. The EDSGC has been put together starting from survey plate scans performed in Edinburgh using the COSMOS machine (Beard et al. 1990), which provides a relative positional accuracy $\sim 0.5$ arcsec (over small areas). This is ideally suited for use with an optical fibre spetrograph, where light is collected over apertures of 1-2 arcseconds diameter (see next section). In addition, the faint magnitude reached by these catalogues allows to easily match the density of fibres on the spectrograph. In the case of the 33 arcmin field of Optopus, the fibre spectrograph used here, the match with the average galaxy number counts in the $b_J$ band is for $b_J \simeq 19.2 - 19.5$.

As discussed in the introduction, the observations presented here were mostly intended as preliminary tests of the performances of the upgraded version of the ESO fibre spectrograph Optopus when used for medium-deep redshift survey work. We already had at our disposal from previous observations $\sim 170$ unpublished redshifts in the area of the cluster Klemola 44 (K44 hereafter), near the South Galactic Pole. Our effort concentrated in trying to collect a large quantity of radial velocities for galaxies in this same area, that we named PL, to allow a very detailed kinematical study of the cluster. This area was covered with a mosaic of 18 Optopus fields, some of which had to be observed twice given the large number of objects with respect to the average number counts. The honeycomb geometry of the mosaic is shown in fig. 1a. In addition to this region, we observed another two smaller areas east and west of K44, called respectively PW and FD. These were meant to provide clustering information in depth in areas not dominated by a single conspicuous cluster, with the practical benefit of allowing a more homogeneous right ascension coverage during the night. Figures 1b and 1c show the disposition of the Optopus fields in these areas. A fourth region, PE, was also partly covered during the observations (fig. 1d), but due to bad weather conditions only a few redshifts in two Optopus fields were obtained. These are also presented here, but are evidently too scarce for allowing any possible use of the PE region in the parallel analyses we are performing (Ettori et al. 1995a, 1995b). Table 1 gives the central coordinates of the single Optopus fields observed. Fig. 2 gives an all-sky view of the positions of the observed fields, indicating also for comparison the region covered by the EDSGC, the position of the famous NGP-SGP pencil-beam of the faint galaxy survey of Broadhurst et al. (1990 – BEKS), and the area covered by the ESP survey.

### 2.2 Observations

Optopus is a fibre-optic coupling system developed for the Cassegrain focus of the ESO 3.6 m telescope, which allows multi-object spectroscopy within a field of 33 arcmin diameter. Alluminum starplates with fibre holders have to be preparared in advance of the observation. These are eventually mounted at the telescope and the fibres are manually inserted into the apertures. At their opposite end, the fibres form the entrance slit of the spectrograph. In its original version, before 1988, the system suffered from poor throughput of the fibres, mainly due to the presence in front of each fibre of a microlense used to convert the focal ratio from the $f/8$ output of the Cassegrain focus to $f/3$ (Lund & Enard 1984). In the new version, which was tested for the first time during our observations, new Polymicro fibres with reduced focal-ratio degradation and enhanced blue-UV transmission, are placed directly in the focal plane of the telescope. The new fibres allow direct coupling to the $f/8$ collimator of a *Boller & Chivens* spectrograph. The global improvement in efficiency of the new configuration has been evaluated to be about one magnitude (Avila et al. 1989).

For our program, the ESO grating no. 15 was used together with the $f/8$ collimator and $f/1.9$ blue camera of the B&C spectrograph. This configuration gives a dispersion of 170 Å/mm and a FWHM resolution of about 10 Å. For studies of the large-scale distribution of galaxies it is necessary to keep the error on the measured redshifts below 100 km s$^{-1}$. The use of the cross-correlation technique for the redshift estimate allows *rms* errors between 1/25 and 1/10 of the nominal, single-line resolution, to be reached, depending on the S/N ratio of the spectrum. The expected uncertainty on $cz$, given our spectroscopic setup, was then between 25 and 60 km s$^{-1}$. As we shall see, this is well matched by the actual errors estimated on the measurements.

The observations were performed during 7 nights in August/September 1988 and 4 nights in October 1989. During the 1988 run, 2.5 nights were lost due to bad wheather; in 4.5 nights of actual observations, 21 independent Optopus fields were observed. The success rate related to bad weather or technical problems, for the 1989 run was around 80%, with 17 fields observed. Some of these were re-observations of some of the PL fields, to increase their completeness. The total number of independent fields in the whole survey on which at least one redshift was obtained is 30, as also shown by fig. 1.

Total exposure times were between 60 and 90 minutes, typically divided into two–three exposures aimed at allowing accurate cosmic–ray elimination. Three to four fibres were used during each exposure to collect the sky spectrum. He–Ar calibration spectra were observed immediately before and after the science exposures, through the same fibre configuration and with the telescope in the same position, as well as quick lamp flat–field frames. Standard bias frames were taken in multiple exposures at the beginning and end of each night.

## 3 DATA REDUCTION AND REDSHIFT DETERMINATION

### 3.1 Standard Reduction and Wavelength Calibration

All the data were reduced using the IRAF package running on a SUN SparcStation II at the Osservatorio Astronomico di Brera. After properly subtracting the bias – as obtained from several, median averaged, single bias exposures – and trimming the outer superfluous regions of the CCD frames, the 2–3 exposures available for each field were combined. This was performed in general through a k–$\sigma$–clipping algorithm intended for eliminating cosmic ray (CR) events. The proper IRAF task was then used to extract the single spectra from the frames. The algorithm performs an optimal extraction, following also possible distortions of the spectra along the dispersion direction. It also allows rejection of discrepant pixels, as residual CR events which survived the initial cleaning phase. The extraction model profile was obtained from the high–S/N–ratio flat–field of each frame, and contemporarily to the science spectra, the corresponding He-Ar spectra were also extracted. The final result of this operation was represented by two sets of $n \leq 31$ one–dimensional spectra (objects plus arcs), for each observed field, ready for the subsequent wavelength calibration. To improve the stability of the wavelength calibration solution, in addition to the He-Ar lines we used the strong night–sky line O I $\lambda 5577$ Å . The position of this sky line is particularly well suited to fill the gap existing in the He-Ar spectrum between 5100 and 5800 Å, where no bright enough line is present. The addition of the O I $\lambda 5577$ Å line allowed us to work with a nice set of $\sim 10$ very good S/N lines, evenly distributed over the spectral range of interest, and obtain typical $rms$ calibration errors $\sim 0.2$ Å.

### 3.2 Sky Subtraction

Sky subtraction for optical fibre data is typically not a straightforward task (see e.g. Ellis et al. 1984). In our case, we adopted the following strategy. During the observations, a typical number of 4 fibres in each field were dedicated to collecting light from the night sky. The position of these fibres was fixed during the starplate preparation phase, and chosen to avoid cross-talk with the object fibres. As we shall see, in a few cases a sky fibre was found to have collected light from an unexpected object, evidently fainter than the limit of the survey at $b_J = 19.5$, providing a redshift for an additional 'foreign' galaxy.

The main issue for achieving a sufficiently accurate sky subtraction is the normalization of the throughput of the single fibres to a common scale. The throughput is essentially independent of the wavelength, but is strongly related to the fibre bending and torsion, thus to the actual position of the telescope. One possibility for normalizing the relative transmission of the sky and object fibres is to use the continuum lamp flat–fields observed before and after each science exposure. However, we tried to use an even more direct method to measure the actual throughput based on the relative fluxes in the bright night–sky lines. We first developed an IRAF procedure which automatically measures the flux $f$ in the two sky emission lines O I $\lambda 5577$ Å and O I $\lambda 6300$ Å for all the spectra of the same Optopus field. The program measures the ratio

$$r_f = \frac{f(\text{OI}\lambda 5577\text{Å})}{f(\text{OI}\lambda 6300\text{Å})} \quad ,$$

and then calculates the mean and standard deviations of the distribution of $r_f$ among the different spectra. The aim of this was to identify spectra for which the estimate of the flux in one or both the sky lines used is corrupted by, e.g., the coincidence with a galaxy absorption or emission feature or with a residual CR spike. We defined therefore as 'good' those spectra with $r_f$ values within $\langle r_f \rangle \pm \sigma_{r_f}$. Among these we chose as reference the spectrum with the highest O I $\lambda 5577$ Å line flux, and normalized each spectrum multiplying it by the ratio of its O I $\lambda 5577$ Å line flux to the reference one. Spectra which had $r_f$ outside of the $\pm 3\sigma$ interval around the mean were flagged and inspected. Typically, one of the two sky lines was affected by a feature in the galaxy spectrum. In this case, we used for the normalization the remaining 'good' line. The result of this operation was thus a set of 31 spectra, normalized to a common response curve, among which 3-4 were pure sky spectra.

The sky spectra were then averaged through a 3-$\sigma$-clipping average, constructing a mean sky spectrum which was subtracted from all object spectra belonging to the same field. The results were individually checked, and residuals from not perfectly subtracted bright sky lines were manually eliminated by interpolation of the adjacent continuum. A few examples of spectra of different quality are shown in fig.3.

### 3.3 Spectral Templates

The basis of the cross–correlation technique for redshift measurement is the comparison of the object spectrum with a model spectrum of known radial velocity and ideally infinite S/N ratio, the *template*. The power of the technique lies in the remarkable similarity in the basic features among galaxy spectra, although the relative intensity of absorption lines can vary quite significantly, in particular when different morphological types are considered. In practice, to cover the range of spectral properties a number of different templates is used for each object and the one producing the highest cross–correlation peak is then taken to be the best model, at zero radial velocity, of the galaxy spectrum being measured.

*Internal Stellar Templates*

In the present work particular care has been devoted to the definition, choice and – in some cases – construction of the template spectra. After the first set of observations was performed, it was clear that some of the stars misclassified as galaxies in the EDSGC ($\sim 5\%$) and thus observed during the survey, were quite appropriate for being used as stellar templates. This is the case for late $G$ and $K$ spectral types, i.e. those which represent the dominant contribution to the global galaxy spectrum. Having also at our disposal a library of galaxy and further stellar templates used for other projects (see below), no other template was specifically observed during the subsequent runs of the survey.

It is instructive to detail the procedure followed to select the best 'internal' templates (i.e. observed with the same instrumentation as the objects to be measured). We first inspected directly all the brightest stellar spectra among the available data, checking the prominence of the H$_\beta$ $\lambda 4861$ Å, Mg I $\lambda 5175$ Å, Na I $\lambda 5892$ Å lines, the G $\lambda 4304$ Å and Ca+Fe $\lambda 5269$ Å bands. In this way we selected a group of 60 $G - K$ stars, potentially suited for being used as templates. During the visual inspection of the spectra, we qualitatively divided them into 5 broad spectroscopic subgroups on the basis of similar intensity ratios among the above mentioned absorption features. This allowed us to construct 5 synthetic spectra by averaging, after weighting for the continuum intensity, the spectra in the five classes. Note that the spectra were combined without correcting their own heliocentric radial velocity, in an attempt to mimick the broadening effect produced by the stellar velocity dispersion on the galaxy spectrum absorption lines. Obviously, this is quite an arbitrary operation, and cannot be considered as a method to properly construct a synthetic galaxy spectrum starting from its basic stellar components. The only aim is to empirically improve the significance of the cross–correlation output, to be checked *a posteriori* on the actual results. In fact, the cross–correlation of the single spectra of each group among themselves showed a typical dispersion among the different stars $\sim 50$ km s$^{-1}$, i.e. of the order of the errors on each single cross–correlation measurement, but still smaller than typical internal velocity dispersions in galaxies. A direct cross–correlation test of the synthetic spectra on a set of science spectra showed that these were performing in a number of cases better, i.e. providing a more significant peak, than the single components.

The performances of the 60 single stellar template candidates were tested by applying them (after determining their zero–point velocity as described below), to estimate the redshift of three galaxies with accurately known radial velocity, for which we had high S/N spectra from a previous project (Collins et al. 1995): NGC 6070, NGC 5746 and NGC 5796. All these galaxies have 21 cm line redshift measurements with error $< 10$ km s$^{-1}$ (see discussion on Table 3 in the following). As a result, we selected the 4 best stellar templates on the basis of their ability of approaching the 'true' redshift of the three galaxies with high significance.

In summary, at the end of this selection procedure we had at our disposal 9 'stellar' templates observed with the same identical instrumentation as the target galaxies of the survey, 4 single stars plus 5 synthetically combined spectra.

*Internal Galaxy Templates*

After running a first test cross-correlation session over the whole set of spectra – as described below – we realized how some spectra of galaxies in the survey with particularly good S/N ratio were consistently giving very significant cross–correlation peaks, with a confidence parameter (see below for definition), $R \sim 10$ or larger. In addition, we noted how the ability of the cross–correlation algorithm to efficiently detect redshifts larger than 10000 km s$^{-1}$ was clearly reduced by the fact that internal stellar templates have a starting wavelength $\sim 4200$ Å. In this way, important absorption features at lower wavelength, especially the Ca K $\lambda 3934$ Å and Ca H $\lambda 3968$ Å lines, are not present in the template spectrum. These absorption lines enter the observed wavelength range for objects with $cz > 20000$ km s$^{-1}$, and are therefore very useful for estimating the redshift of distant galaxies with typically faint spectra. We found therefore particularly important to include in the final set of templates a number of galaxies with high recession velocity, selected among those giving high values of $R$. In the attempt to cover a redshift range as wide as possible, we assembled the best galaxies of the survey, defined as having a best confidence parameter $R > 10$, into four broad distance classes, characterized respectively by $cz \sim 10000$, $\sim 20000$, $\sim 30000$ and $> 30000$ km s$^{-1}$. For each class, the object with the largest average $R$ among the different templates was promoted as a new galaxy template. The average was performed excluding the best and the worst of the values of $R$ obtained with the different templates, to enhance those spectra with the more general spectral features. The second best $R$ galaxy in the first class was also included in the selected set due to its high S/N ratio and the particularly good performances with all the templates.

In summary, this selection provided us with 5 further templates observed with the same identical instrumentation as the survey objects. The important feature of this new set is that they are galaxy spectra, with good blue coverage.

*Template Zero Point*

The above described operations produced a set of 14 internal templates: 4 stars, 5 composite stars and 5 galaxies. A crucial operation for minimizing the error on the redshift measurement is the determination of the radial velocity for each template, i.e. what will represent the zero points of our velocity measurements. To this end, we adopted a two–step procedure which can be described as follows. Andrew Connelly kindly provided us with a set of 7 radial velocity Henry Draper (HD) standard stars, observed at the AAT telescope with moderately high resolution (1.4 Å/px) and high S/N ratio, covering the wavelength range 3500−6500 Å. For these stars, a very accurate radial velocity is provided in the Astronomical Almanac, as detailed in Table 2. Our final goal was to use these (or some of these) higher precision templates as zero–point velocity calibrators for our internal set of templates. To this end, we had to be absolutely sure that the actual radial velocities of the HD spectra were consistent within their errors with those quoted in the Almanac. These spectra had nearly five times higher resolution than those in our survey and could therefore provide a very tight constraint on the redshift of our internal templates (obviously within the intrinsic resolution of the latter ones). The only uncertainty about the HD radial velocities was the possibility of some systematic error in the spectra available to us, which could have introduced non–negligible shifts with respect to the published values. To ascertain this, we adopted a method which differs slightly from that applied by Tonry & Davis (1979; TD79 hereafter), but which proved to be quite was to cross–correlate the HD templates one against each other, assuming as correct radial velocities the literature values. In fig. 4a we plot the result of the cross–correlation , showing for each HD spectrum $i$ the value of the quantity

$$\Delta v_{ij} = v_{ij} - v_i^{lit} \quad ,$$

i.e. the difference of the velocity of $i$ measured by $j$ and its literature value. Each point in the plot is characterized by the value of $j$ indicating the template which produced that value for the spectrum $i$ reported on the abscissa. The underlying hypothesis is that if only random errors are present, the $\Delta v_{ij}$ should be nearly normally distributed around zero, with a dispersion of the order of that expected from the resolution of the HD spectra, $\sigma \sim 25$ km s$^{-1}$. It is clear from the figure that this is not the case for some of the objects analyzed. Qualitatively, the barycenter of the distributions of the $\Delta v_{ij}$ is significantly different from zero for about half of the spectra, and the distributions are not fully symmetric. The global indication of the figure is that some of the HD spectra do have some some significant shift with respect to their literature redshift. In particular, it is clear the tendency of #8 to overestimate the redshifts of the other spectra, while the measurements using #1 are systematically underestimated. Given the small number statistics and the possible asymmetry of the distributions, it is clear that the mean value of the $\Delta v_{ij}$ for each $i$ cannot be considered as a robust estimator for the true zero–point velocity. A more sensible choice seems to be the median, which is not affected by the asymmetric tails produced by the presence of a small number (1–2 in our case) of very discrepant templates (like, e.g., #8). We therefore determined a new value of $v_i^{lit}$, by taking the median of the seven $\Delta v_{ij}$ for each $j$. With these new values for each spectrum, we can construct new distributions of the $\Delta v_{ij}$, re-determine the median and obtain new $v_i^{lit}$. The procedure can be repeated a number of times, but essentially converges already after three iterations. In fig. 4b we show the same plot of fig. 4a, after the third iteration. Note how the global scatter has been significantly reduced with respect to the first iteration, and is now compatible with that expected from the intrinsic resolution of the spectra. Essentially, the zero–point changed significantly only for three templates, as can be seen from Table 2. For the remaining spectra, the difference is less than 15 km s$^{-1}$, i.e. within the expected intrinsic uncertainty. These five spectra represent therefore a carefully tested set of templates with known radial velocity.

The idea is now to use these higher precision, well calibrated templates to calibrate in turn the zero point of the internal templates previously prepared. To this end, we ran the cross–correlation over the internal templates, using the five best HD as templates. For each internal spectrum, the velocity obtained with the best $R$ value was chosen as zero point (i.e. its 'literature' redshift), after checking the global consistency of the values obtained by the five templates. It is reassuring that the agreement among the five templates was very good, with typical $rms$ discrepancies around 10–15 km s$^{-1}$. Table 3 reports the final radial velocities for the set of internal templates, together with the $rms$ value of the differences between the five measurements and their mean. In the same table we list also the three high S/N galaxies from Collins et al. (1995), which were included among the used templates to provide three external reference galaxies. These proved to be useful – despite a slightly lower intrinsic resolution – for corroborating the significance of uncertain redshifts for which the unanimity of the internal templates could be seen as suspect. In addition, we included in the final set of templates used for the cross–correlation with the survey spectra the two best performing HD stars, HD02 and HD07, for a total of 19 spectra.

### 3.4 Redshift Measurement

As we have seen, the redshifts were measured using the cross–correlation technique discussed in detail by TD79, in the version developed at the Smithsonian Astrophysical Observatory as a subpackage of IRAF – called RVSAO – which closely follows the original prescription. Briefly, the cross–correlation of the observed galaxy spectrum with the template spectrum is accomplished by taking the Fast Fourier Transform (FFT) of the two spectra, multiplying the two FFTs, and then transforming back the result to get the Cross-Correlation Function (CCF). The highest peak of the CCF is related to the radial velocity difference between the two spectra. Before actually starting this machinery, the two spectra are rebinned into logarithmic bins, so that the relative redshift becomes a linear shift between the two, and a number of operations are performed on them to improve the quality of the final result. These are described in detail by TD79, and include continuum subtraction, cosine-bell apodizing and bandpass filtering. RVSAO provides a user–friendly interface and the possibility to check all the intermediate steps, as well as to try several experiments to find the best set of parameters. Specifically, the spectra were rebinned into 4096 logarithmic bins, which corresponded to a formal velocity binwidth of $\sim 35$ km s$^{-1}$. The spectra were then filtered in order to eliminate both the low frequency spurious components left by the subtracted continuum, and the high frequency binning noise. The best set of filter parameters was chosen to maximize the significance of the CCF. The peak of the CCF is fit by a quadratic polynomial, determining both the wavelength shift – from its position – and an estimate of the error $\varepsilon$ – from its width. In output, together with the value of $cz$, the program provides the value of $\varepsilon$, of the peak height, and of the confidence parameter $R$, that, as defined by TD79, corresponds to the ratio between the height of the highest peak in the CCF and the average value of the adjacent background.

After a first test run, the cross–correlation routine was run over the whole set of available spectra, including about 1000 spectra. The results obtained were scrutinized one by one, and ambiguous cases were re-examined directly. Spectra with $R < 3$ were typically discarded. The result was considered significant only when at least half of the templates indicated a similar velocity, with the further constrain that these should not be templates of the same origin (e.g. all internal stars). We found, indeed, a tendency of 'similar' templates to be sometimes biased towards similar values, certainly because of some specific instrumental features not removed during the continuum subtraction and filtering phases. This is particularly true when $R \sim 3$, i.e. for very low S/N spectra. In general, measurements with $R < 4$, although included in the final list, have to be regarded with

## 4 THE DATA

The final galaxy data are presented in Tables 4a–d, subdivided among the four different regions. The column are self-explanatory. RA, DEC, and $b_j$ are from the EDSGC, velocities are in km/sec. The total number of newly observed radial velocities which were accepted according to the above given criteria is 602, including 494 galaxies and 108 stars. However, due to a mismatch between the fibre and spectral numbers, 7 spectra in the PL region do not have a certain identification (and thus RA and DEC), and are excluded from the tables. In Table 4e we have also listed the redshifts for those galaxies within the PL region which already had an (unpublished) measurement from previous observations by one of us (MT). These are reported here for completeness, since they were not re-observed in the present survey, and will be referred to in the following as the MT sample. Earlier, less precise coordinates for these objects (originally selected from deep ESO 3.6 m telescope prime-focus plates), were matched to the EDSGC, and when possible a more precise position has been found, together with the corresponding $b_j$ magnitude. Galaxies which did not appear in the EDSGC are indicated; no magnitude is given for these objects. The redshift errors for these external data are $< 100\ km\ s^{-1}$, thus comparable to those of the new data. Details will be presented in a parallel paper which concentrates on the K44 region (Ettori et al. 1995b).

The completeness of the data, defined at the given magnitude limit as the ratio of the number of galaxies with measured redshift over the total number of EDSGC galaxies, is shown in Table 5. The numbers quoted have been cleaned from those objects which were shown to be stars by the spectroscopy, and were thus deleted from the parent galaxy catalogue. For the PL region, the total sample selected from the EDSGC (809 galaxies), is explicitly split into the number of objects considered for the present survey (748) plus those from previous observations (61).

The redshifts listed as *noedsgc* pertain to objects which: 1) have been serendipitously observed through one of the 'sky' fibres, on positions where no EDSGC (or previously visually identified) galaxy is present; 2) (only for PL) are part of the MT sample, which was selected visually from deep ESO-3.6m telescope prime-focus plates, and are missing from the EDSGC (17 objects overall the PL region).

In general, the incompleteness is mostly due to objects which produced a too low S/N spectrum, i.e. either had too low a surface brightness or were observed through a bad fibre or in bad wheather/seeing conditions. Very few objects, in fact, were not actually observed. Note also that, since the sample was selected from an early, unofficial version of the EDSGC, after the final photometric recalibration the actual magnitude limit of the PL region turned out to be $b_j = 19.62$ and not 19.5. Therefore, in Table 5 the global completeness for PL5 is referred to this slightly fainter limit, this anomaly being indicated by the asterisk in the column of the $\leq 19.5$ completeness. The effects of the incompleteness on the specific analyses which have been carried out on the present data are discussed in more detail in the relevant papers.

It is interesting to plot the distribution of the $R$ parameter and of the velocity errors $\varepsilon$ among the whole set of measurements. These are shown in fig. 5, and give an indication of the overall quality of the redshifts. Fig. 5a is constructed using the whole set of measurements, i.e. including also those spectra which were not good enough to produce a significant estimate of the redshift. The histogram is divided into two parts by the dashed line showing the value of $R$ chosen as the borderline below which redshifts were generally discarded. Fig. 5b gives the distribution of the RVSAO errors $\varepsilon$ for those estimates with $R > 3$, showing how the median value of $\varepsilon$ for the data presented here is $\sim 75\ km\ s^{-1}$.

The error provided by the cross-correlation package, $\varepsilon$, is an internal error, and is estimated on the basis of some assumptions which have been recently criticized by Heavens (1993), who proposed an alternative method for constructing a more significant error estimate. The algorithm we used is based on the original definition by TD79. We can obtain an independent estimate of the external errors by using those galaxies for which two independent observations were obtained in the survey. We have 36 pairs for which both spectra provided an acceptable redshift. In fig. 6 we plot the distribution of the absolute differences of the two values. The median of this distribution is $100\ km\ s^{-1}$, indicated by the arrow, i.e. about 1.5 times that obtained from the internal estimate of fig. 5b.

A pictorial representation of the large-scale distribution along the observed beams is given in figures 7 and 8. Fig. 7 shows a global wedge diagram of the data, while in fig. 8 we plot the histograms of the redshift distribution in the observed fields (PE is excluded). In PL, note the expected concentration of galaxies around K44 ($cz_{K44} \sim 9000\ km\ s^{-1}$), but also an indication for further peaks with $\sim 100\ h^{-1}\ Mpc$ separation. The same suggestion seems to arise when examining PW. These qualitative considerations will be extended with a more quantitative analysis of the possible presence of a preferred clustering scale, as suggested by BEKS, in a forthcoming paper (Ettori et al. 1995a).

**Acknowledgments**

We are indebted to G. Avila for beautifully designing the upgraded version of Optopus and for his help during the starplate preparation and the observations. This effort would have not been possible without the generous contribution of C.A. Collins, who provided us with galaxy positions and magnitudes from the EDSGC in advance of publication, and the help of N. Heydon-Dumbleton. Their neverending enthusiasm and patience is gratefully acknowledged. We thank G. Chincarini for many useful discussions, A. Connelly for providing us with the HD templates and R. Merighi for his important help about the use of the cross-correlation package RVSAO.

**REFERENCES**

Avila, G., D'Odorico, S., Tarenghi, M., & Guzzo, L., 1989, *The Messenger*, 55, 62

Beard, S.M., MacGillivray, H.M. & Thanisch, P.F., 1990, MNRAS, 247, 311

Broadhurst, T.J., Ellis, R.S., & Shanks, T., 1988, MNRAS, 235, 827

Broadhurst, T.J., Ellis, R.S., Koo, D.C., & Szalay, A.S., 1990, *Nature*, 343, 726 (BEKS)

Collins, C.A., Guzzo, L., Nichol, R.C. & Lumsden, S.L., 1995, MNRAS, in press

Ellis, R.S., Gray, P.M., Carter, D., & Godwin, J., 1984, MNRAS, 206, 285

Ellis, R.S., & Parry, I.R., 1988, in *Instrumentation for Ground Based Optical Astronomy*, L. Robinson ed., New York: Springer, p.192

Ettori, S., Guzzo, L., & Tarenghi, M., 1995a, in preparation

Ettori, S., Guzzo, L., & Tarenghi, M., 1995b, in preparation

Geller, M.J., & Huchra, J.P., 1989, Science, 246, 897

Giovanelli, R., Haynes, M.P. & Chincarini, G., 1986, ApJ, 300, 77

Giovanelli, R., & Haynes, M.P., 1991, *Ann. Rev. Astr. & Astroph.*, 29, 499

Gray, P.M., 1984, in *Proc. SPIE Instrumentation in Astronomy, V, 445*, A. Boksenberg & D.L. Crawford eds., p.57

Guzzo, L., & Tarenghi, M., 1987: ESO Internal Report, October 1987

Heavens, A.F., 1993, MNRAS, 263, 735

Heydon-Dumbleton, N.H., Collins, C.A., & MacGillivray, H.T., 1989, MNRAS, 238, 379

Hill, J.M., 1988, in *Fiber Optics in Astronomy*, S.C. Barden ed., San Francisco: PASP, p.77

Huchra, J., Davis, M., Latham, D., & Tonry, J., 1983, ApJS, 52, 89

Klemola, A.R., 1969, ApJ, 74, 80

Lund, G., & Enard, D., 1984, in *Proc. SPIE Instrumentation in Astronomy, V, 445*, A. Boksenberg & D.L. Crawford eds., p.65

Maddox, S.J., Efstathiou, G.P., & Sutherland, W., 1990, MNRAS, 246, 433

Oort, J.H., *Ann. Rev. Astr. & Astroph.*, 21, 373

Rood, H.J., *Ann. Rev. Astr. & Astroph.*, 26, 245

Tonry, J., & Davis, M., 1979, AJ, 84, 1511 (TD79)

G. Vettolani, E. Zucca, A. Cappi, R. Merighi, M. Mignoli, G.M. Stirpe, G. Zamorani, C. Collins, H. MacGillivray, C. Balkowski, J.M. Alimi, A. Blanchard, V. Cayatte, P. Felenbok, S. Maurogordato, D. Proust, G. Chincarini, L. Guzzo, D. Maccagni, R. Scaramella & M. Ramella, 1994, in *Studying the Universe with Clusters of Galaxies*, H. Böhringer & S. Schindler eds., Garching: MPE report n.256, p. 7

Zwicky, F., Herzog, E., Karpowicz, M., Kowal, C.T., Wild, P., 1963-68, *Catalogue of Galaxies and Clusters of Galaxies*, Pasadena: Cal. Inst. Technol., 6 Vols

**Figure Captions**

**Figure 1 (a,b,c,d).** Geometrical configuration of the observed 33 arcmin Optopus fields. With the exception of the FD region, the galaxies were selected to lay within adjacent exagon constructed within each Optopus field. This honeycomb geometry is the simplest way to fully cover an area of the sky using a circular field instrument.

**Figure 2.** Aitoff all-sky projection showing the position of the observed fields on the celestial sphere, compared to that of the BEKS SGP beam (open circle). The large rectangular region shows the area covered by the EDSGC (60 Schmidt plates), while the thin East–West strip is the area surveyed by the ESP project of Vettolani et al. (1994).

**Figure 3.** Some examples of spectra of different quality from the survey. Spectra are not flux-calibrated, thus the ordinates give just relative intensity in arbitrary units. a) Spectrum of a $b_j = 15.90$ galaxy (RA=23:41:34.1, DEC=-28:34:35), which gave a redshift with very high confidence, $R = 19.3$. b) Example of spectrum giving still a good confidence level ($R = 5.13$. RA=23:44:27.1, DEC=-28:14:08, $b_j$=18.70), although having possibly had some problems in the sky subtraction. c) Example of a lower $S/N$ spectrum (RA=02:14:35.3, DEC=-28:30:48, $b_j$=18.91), with $R = 3.90$, which is the median value of the distribution of $R$ among the whole set of spectra. d) Example of a very low S/N spectrum (RA=23:43:04.1, DEC=-28:23:45, $b_j = 19.36$), which gave an $R = 1.87$, the lowest value of $R$ among the data included in the final table. Cases like this, i.e. below the exclusion threshold of $R = 3$, are passed only after positive visual inspection of the cross-correlation function. They have in any case to be considered as tentative.

**Figure 4 (a,b).** Results of the cross-correlation among the high-quality HD spectra, showing for each spectrum $i$ the value of the difference of the velocity of $i$ as measured by $j$ and its literature value. Each point in the plot is characterized by the value of $j$ indicating the template which produced that value for the spectrum $i$ reported on the abscissa (see text for details).

**Figure 5.** a) Distribution of the value of the confidence parameter $R$ among the whole set of measurements (including also those spectra which did not produce a significant redshift). The dashed line shows the value of the threshold in $R$ which has been chosen as the gross borderline below which redshifts were generally discarded. b) Distribution of the RVSAO errors $\varepsilon$ for the selected spectra (i.e. with $R > 3$); the median value of $\varepsilon$ is $\sim 75\ km\ s^{-1}$.

**Figure 6.** Distribution of the absolute differences between repeated observations of the same galaxy. The median of the distribution is 100 $km\ s^{-1}$ (arrow), i.e. about 1.5 times that obtained from the internal estimate of fig. 5b.

**Figure 7.** Right Ascension wedge diagram showing the global distribution of the galaxies along the directions surveyed.

**Figure 8 (a,b,c).** Redshift distribution along the three main directions explored in our study. A tendency to some regularity in the spacing among the peaks of the histogram seems to be evident in the case of PL and PW.